# UAV Active Perception and Motion Control for Improving Navigation Using Low-Cost Sensors

Names: Konstantinos Gounis , Nikolaos Passalis , *Member, IEEE* and Anastasios Tefas , *Member, IEEE*

*Abstract*- In this study a model pipeline is proposed that combines computer vision with control-theoretic methods and utilizes low – cost sensors. The proposed work enables perception-aware motion control for a quadrotor UAV to detect and navigate to objects of interest such as wind turbines and electric towers. The distance to the object of interest was estimated utilizing RGB as the primary sensory input. For the needs of the study, the *Microsoft AirSim* simulator was used. As a first step, a *YOLOv8* model was integrated providing the basic position setpoints towards the detection. From the *YOLOv8* inference, a target yaw angle was derived. The subsequent algorithms, combining performant in computational terms computer vision methods and *YOLOv8*, actively drove the drone to measure the height of the detection. Based on the height, an estimate of the depth was retrieved. In addition to this step, a convolutional neural network was developed, namely *ActvePerceptionNet* aiming at active *YOLOv8* inference. The latter was validated for wind turbines where the rotational motion of the propeller was found to affect object confidence in a near periodical fashion. The results of the simulation experiments conducted in this study showed efficient object height and distance estimation and effective localization.

## I. INTRODUCTION

UAVs, also known as drones are being extensively considerable candidates for civil and commercial applications [1]. In parallel, deep learning methods show outstanding results for solving a wide variety of robotic tasks in the areas of vision, perception, planning, localization, and control creating a lot of excitement in the research community [2] [3].

The development of realistic environments that enable the simulation of the underlying UAV sensor physics, optics / electromagnetics and the vehicle's rigid body dynamics whilst the UAV is represented by its digital 3D counterpart [4] is of a particular importance. *Microsoft AirSim* is a free, open-source, cross-platform simulator for vehicles, built on *Unreal Engine*.

*AirSim* provides application interfaces that can retrieve sensory data or send control signals to the vehicles under test through. The UAV dynamics, following the Newton-Euler equations of motion, are simulated by a dedicated high-fidelity physics engine [5].

*YOLO* series of deep convolutional network models have marked a great success in the field of computer vision. Compared to its successful predecessors *YOLOv8* is an advanced and cutting-edge model that offers higher detection accuracy and speed [6]. However, the integration of big deep models in UAVs with limited resources is still challenging, as the latencies accrued from the deep model inference can even lead to stability issues in control.

In the UAV visual navigation pipeline, it is crucial to address other important aspects related to the perception and efficient tracking of the image features of interest. Several studies have been conducted on tracking image feature points for single or multi-object tracking within subsequent video frames, such as [7], [8], [9], and [10]. On the other hand, several studies have approached the tracking problem from the perspective of relative motion kinematics, illustrating the close relationship between the fields of vision, perception, and control in the context of perception-aware UAVs. Amongst them are optical flow-based velocity estimation and control [11], active structure-from-motion [12], active perception control [13], [14] and learning-based perception [15].

According to Bajcsy et.al. [16] an observer is considered active when it is engaged in any activity whose purpose is to control the geometric parameters of the sensory system in order to achieve a sensing / perception objective. In the same paper it is stated that an actively perceiving agent is characterized by the ability of determination of the "Why" of its behavior and then attempts to answer at least one of the "What", "How", "Where" and "When" questions throughout the control of each behavior it exhibits. Back in 1988, Bajcsy [17] in her paper defined the active perception as study of modeling and control strategies for perception. The modeling strategies are both local model-based and global model-based. The former focuses on aspects of the sensor model such as optical distortions of the lens, focal lens,

This paragraph of the first footnote will contain the date on which you submitted your paper for review, which is populated by IEEE. It is IEEE style to display support information, including sponsor and financial support acknowledgment, here and not in an acknowledgment section at the end of the article. This work was partially supported by the RoboSAPIENS project funded by the European Commission's Horizon Europe programme under grant agreement number 101133807. This publication reflects the authors' views only. The European Commission is not responsible for any use that may be made of the information it contains.*(Corresponding author: Konstantinos N. Gounis).*

Konstantinos Gounis is with the School of Informatics, Faculty of Sciences, Aristotle University of Thessaloniki, Thessaloniki GR 54124 Greece (e-mail: konstgouni@ csd.auth.gr).

Nikolaos Passalis is with the School of Informatics, Faculty of Sciences, Aristotle University of Thessaloniki, Thessaloniki GR 54124 Greece (e-mail: passalis@ csd.auth.gr).

Anastasios Tefas is with the School of Informatics, Faculty of Sciences, Aristotle University of Thessaloniki, Thessaloniki GR 54124 Greece (e-mail: tefas@ csd.auth.gr).

spatial resolution and bandpass filter. The latter, on the other hand, focus on the overall system outcome and make predictions on how the individual modules interact. Very recently, Nagami et al. [15] consider active perception as analogous to perception-aware planning and control methods. Based on the aforementioned, the purpose of active perception can be deemed as a wide umbrella to provide the control inputs that compensate for the constraints underlying any observed phenomena of interest, in order to improve the quality of the perceptual results.

Cristofalo [14] developed control algorithms to recursively estimate the 3D location of an object point expressed in camera coordinates. The Extended Kalman Filter (EKF) was implemented, incorporating the state space model of the motion of a point in a moving frame and the perspective projection model as the process and measurement models, respectively. The UAV velocity vector in the camera frame was chosen to be the control input. To improve the way a 3D point of interest was perceived, the process model equations were solved with respect to the sought control vector that minimizes the trace of the posterior covariance matrix, in an online gradient-based optimization scheme.

In the area of motion control for visual navigation, Nagami et. al. [18] proposed a scheme where the full state vector, constituted of position, linear velocity, RPY angles and angular body rates respectively are tracked by first decomposing the problem to a high-level position - linear velocity controller and a lower-level attitude and angular rate controller. An optimal controller based on Hamilton-Bellman-Jacobi equation was used as baseline whereas a Deep Learning based controller was developed upon the optimal controller via Imitation Learning and subsequent control policy learning. The proposed error to drive the angular rate controller was defined in the Special Orthogonal Group $SO(3)$. This error metric has been previously illustrated in the so-called near-globally-stable geometric control for UAVs [19], [20].

In this study, we propose a model pipeline that enhances localization with respect to an object of interest whose height and depth is actively estimated whilst aiming at reducing object confidence related uncertainty as a preliminary step. We consider the scenario where: a) the environment is captured via the RGB sensor with the fine-tuned *YOLOv8* and if a wind turbine or an electric tower is found, its pixel centroid(-s) are localized, b) in the case of wind turbine whose rotational motion has been detected, the *ActivePerceptionNet* predicts the remaining time to the next confidence peak, actively reducing the detection uncertainty, c) the desired yaw angle towards the target is estimated and an active perception-control loop drives the drone towards the detection's world location by first moving on a plane and subsequently moving upwards to retrieve object height, d) by fusing object's world height and YOLOv8 bounding box height, the depth of the detection is estimated and e) the 3D position of the object frontal surface centroid is tracked through the EKF localization module whilst the drone smoothly moves towards it.

The rest of this work is structured as follows. Section II describes the proposed model pipeline. Section III is dedicated to the mathematical and technical background of the developed models and algorithms. Section IV illustrates the results of the simulation experiments and Section V concludes the paper.

## II. PROPOSED SYSTEM OVERVIEW

The overall system architecture is decomposed to the following modules: a) *Object Tracking*, b) *EKF localization* and c) *Planning and Control*.

The primary system measurement sources are the 640x480 RGB image camera, the IMU and the GPS with an integrated altimeter for 3D UAV position measurements. The latter uses as reference the drone home position hence acting as a differential GPS providing metric world position measurements w.r.t (0,0,0).

*Remark 1* (Coincident Coordinate Systems) : The camera coordinate frame was treated as coincident with the quadrotor's inertial measurement frame in this paper. This aligns with the UAV simulation configuration used in this study.

*Assumption 1* (IMU angular body rates and translational body velocity from GPS – IMU fusion) : It was assumed that an estimate of the quadrotor's angular rotation rates, the translational velocities, and Euler angles for the subsequent low-level quadrotor control was available. Linear velocity in the body frame is easily retrieved by fusing the IMU RPY angles through the instantaneous rotation matrix with the GPS velocity in the world frame.

The *Planning and Control* module receives the initial estimate of the detection's 3D position. This estimate is directed to the motion planner that solves offline the timing law problem with a given initial position (home), initial velocity, final position (detection's position estimate) and final velocity, in the three dimensions via cubic polynomials. The entire trajectory is sent to the path following controller that takes over low-level UAV control by converting position feedback control signals to attitude feedback control in a process that will be described in the next section. Thus, the outcome of this module are the thrust force and body axes' torques to drive the UAV towards the goal position.

The *Object Tracking* module, given the initial bounding box by *YOLOv8*, first subtracts all the image regions not enclosed in the bounding box via applying a binary mask and subsequently executes the *Canny* edge detection algorithm to extract the external edges, contours and texture edges of the object of interest. The detected geometry is converted to a list of 2D points that is used to initialize the *Lucas Kanade* optical flow tracker. The latter aims at tracking the features throughout the sequence of images taken every 80 milliseconds as the UAV smoothly flies towards the detection. From the tracked points, the two indicating the object's minimum and maximum pixel coordinates define a bounding box, compatible with the initial object detection setup. The calculated bounding box centroid coordinates, area, ratio and the time derivative of the aforementioned quantities are smoothened through a Linear Kalman Filter. This filter receives the first four of them as measurements and predicts the whole 8x1 bounding box state vector, using a Constant Velocity Model. Finally, the filtered bounding box centroid horizontal and vertical pixel coordinates are directed to the localization module, forming the measurements signal to the EKF.

The *EKF localization* module is developed around a linear time varying process model describing the position of a point with respect to a moving frame and a nonlinear Pinhole camera measurements model. The state vector is the 3D position of the detection's bounding box centroid in the three- dimensional camera frame. The process model is constructed by IMU based and GPS data.

The overall system architecture is illustrated in Fig. 1. The model pipelines for active *YOLOv8* inference as well as height and distance estimation, being additional notes to the system in Fig. 1 are in detail described in the first paragraphs of Section III.

## III. TECHNICAL BACKGROUND

### A. ActivePerceptionNet

The *ActivePerceptionNet* Convolutional Neural Network architecture developed in this study consists of six convolutional layers (*conv2d*) each followed by a Rectified Linear Unit (*ReLU*) activation function layer. The input to the network is the RBG image cropped by the preliminary YOLOv8 predicted bounding box, resized to a 32x32x3 signal. It should be noted that a known, calibrated camera pose in the world frame was considered for this active inference process. The convolutional layers are followed by three fully connected layers. The first two are incorporating the *ReLU* function while the last one is followed by a linear activation function. The output is the scalar quantity representing the remaining time until the YOLOv8 prediction on the rotating wind turbine propeller reaches the next confidence peak. The time-to-next-peak is expressed in seconds. Fig. 2 illustrates *ActivePerceptionNet* as a major part of the active *YOLOv8* inference pipeline.

### B. YOLOv8 model

Let the fine-tuned *YOLOv8* object detection model be represented at a high level of abstraction by $f_{detector}$:

$$\begin{bmatrix} u \\ v \\ w \\ h \end{bmatrix} = f_{detector}(I) \quad (1)$$

$I$ is the RGB image signal, $u$ and $v$ are the detected wind turbine / electric tower bounding box horizontal and vertical pixel coordinates and $w$, $h$ the respective bounding box width and height in pixels.

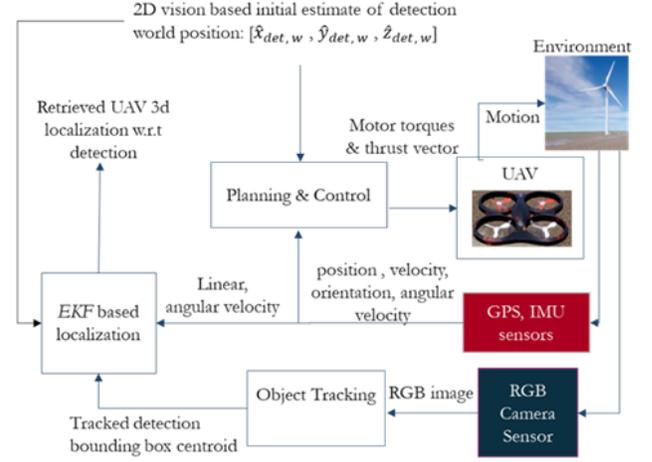

**Fig. 1.** The overall system architecture.

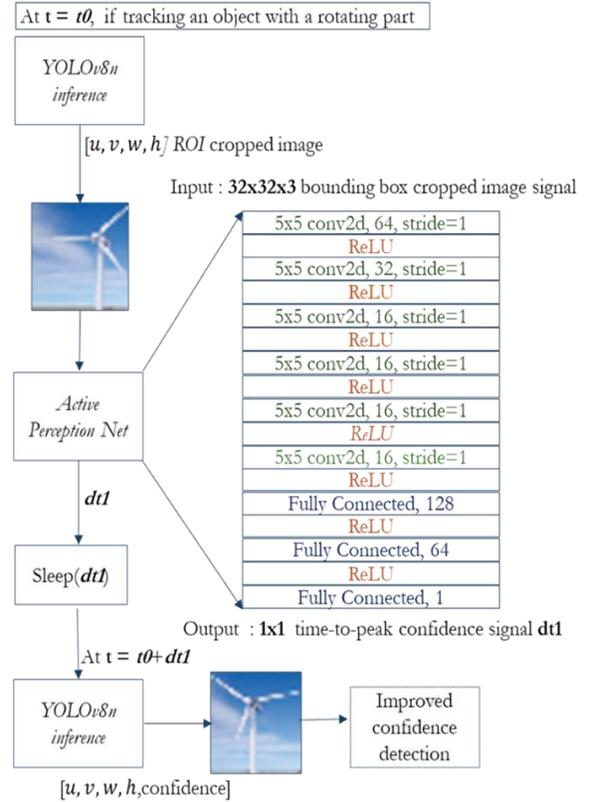

**Fig. 2.** *ActivePerceptionNet* receives the initially inferred YOLOv8 prediction and estimates the system idle time *dt1* to next confidence peak.

### C. Active Height Measurements

The yaw angle, which represents the change in planar orientation required for the UAV to move from the current world position $x_w$, $y_w$ to the world position of a target detected object $x_{t,w}$, $y_{t,w}$ ensuring it directly faces the target during its motion, is defined as:

$$\psi_{des} = tan^{-1}(\frac{y_{t,w} - y_w}{x_{t,w} - x_w}) \qquad (2)$$

We consider a calibrated world position and an orientation described in *SO(3)* by the Identity rotation matrix $I_{3x3}$. Then, the position of a point lying on the frontal surface of the detection, in the camera frame, coincides with the world frame relative distance between the UAV and the detection:

$$\begin{bmatrix} x_c \\ y_c \\ z_c \end{bmatrix} = R^T \begin{bmatrix} x_{t,w} - x_w \\ y_{t,w} - y_w \\ z_{t,w} - z_w \end{bmatrix} \cong I_{3x3} \begin{bmatrix} x_{t,w} - x_w \\ y_{t,w} - y_w \\ z_{t,w} - z_w \end{bmatrix} \qquad (3)$$

$\boldsymbol{x_c} = (x_c, y_c, z_c)$ are the 3D coordinates of the centroid of the object pixel area projected in the camera frame and expressed in meters. It should be noted that in this study, the optical axis aligns with the $x_c$ axis. $R$ is the rotation matrix that converts body/camera coordinates to world coordinates. According to the perspective projection transformation, the centroid coordinates in the image are:

$$\begin{bmatrix} u \\ v \end{bmatrix} = \frac{1}{x_c} \begin{bmatrix} y_c f_l + x_c u_0 \\ z_c f_l + x_c v_0 \end{bmatrix} \qquad (4)$$

$(u_0, v_0)$ are the coordinates of the principal point in the image plane which in this study was the center of the 640 by 480 image plane. $f_l$ is the focal length, considered fixed in this study, which is available via camera intrinsics' calibration. $y_c$ can then be written as:

$$y_c = \frac{(u - u_0) x_c}{f_l} \qquad (5)$$

Plugging (5) and (3) into (2) yields an approximation of the yaw angle of the path to follow:

$$\psi_{des} \cong tan^{-1}(\frac{u - u_0}{f_l}) \qquad (6)$$

Using this information, one observes that the direction to the detection can be preserved. To compensate for the unknown magnitude of the distance to target, a unit meter magnitude was considered, scaled by a factor of 2 at every iteration. Thus, planar position setpoints towards the target were calculated and tracked via the controller described in the subsequent paragraph. The squared pixel ratio of the *YOLOv8* bounding box area to the image was extracted at each setpoint to visually check proximity to the target. A stopping criterion can be defined by the desired portion of the detection in the image. In this study *the rule of thirds* was considered, as in [21]. Fig. 3 illustrates the *Active Height Measurement* model pipeline.

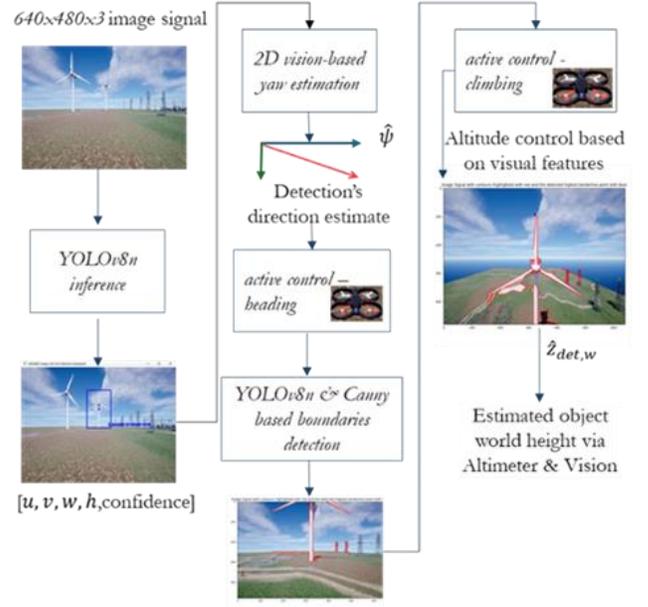

**Fig. 3.** The *Active Height Measurements* pipeline, devising a planar motion until the desired proximity to the target has been achieved and an upward visual features tracking path.

Upon reaching the position where the target is located at the center of the image plane occupying the desired pixel area, an upward motion path is planned. *YOLOv8* was combined with *Canny* edge detection algorithm to identify contours within the bounding box and extract the upper boundaries of the object of interest. In a similar fashion to the planar motion constrained by (6), position setpoints in the world z axis were calculated based on a unit vector iteratively upscaled. The stopping criterion of the upward motion was the vertical alignment of the uppermost contour point which was horizontally closer to the principal point of the image. The latter can be deemed a straightforward computer vision task in the case of a completely stationary target detection with a standard texture such as an electric tower.

The same objective however is not straightforward in the case of a target object that performs motion which in the general case is not necessarily of a constant translational velocity. As in our study amongst the sought objects were wind turbines, the problem of tracking the uppermost point in a rotating propeller to provide a more accurate height estimation was addressed as follows. Given that the UAV had been driven to a considerable altitude from where it is front facing the propeller, the sought point is at the altitude of the highest vertex by the time the propeller is oriented like the 'Mercedes-Benz' logo. In Fig. 3 an example demonstration of the sought propeller orientation is provided.

Let the digital motion detection filter be:

$$|I_t - I_{t-1}|, \qquad (7)$$

where $I_t$ is a thresholded grayscale image signal acquired at time *t* and $I_{t-1}$ the respective signal acquired at the previous time *t-1*. The absolute frame difference in (7) is considered a very performant in computational terms [22] motion segmentation method. To alleviate ghosting effects, the time between two consecutive frame captures was minimized i.e. by capturing and processing the frames one after another. Fig. 4 provides a snapshot of the real-time execution of the filter.

The model that approximated the blade kinematics is:

$$\begin{aligned} \dot{\beta} &= \omega_\beta, \\ \begin{bmatrix} x_p \\ y_p \end{bmatrix} &= \begin{bmatrix} x_{bc} + l_b \sin\beta \\ y_{bc} - l_b \cos\beta \end{bmatrix} \end{aligned} \quad (8)$$

$\beta$ is the blade angle defining its orientation in the image plane and $\omega_\beta$ is the blade angular velocity. $l_b$, $(x_p, y_p)$ and $(x_{bc}, y_{bc})$ are the blade length, the position of the upper-most point and the position of the lower-most point of the blade geometry respectively. Fig. 5 illustrates the blade kinematics in the image plane. All quantities, expressed in pixels, can be easily retrieved with the aid of an off-the-shelf image processing software.

*Remark 2* (Near uniform circular motion) : The bounding box cropped image exhibits periodical changes, as observed through the UAV data acquisition when hovering at a viewpoint. The aforementioned periodicity, caused by the propeller motion, was reflected in the object confidence score. Hence, utilization of such observations revealed the near constant period and angular frequency of the propeller motion.

*Assumption 2* (Coincidence of the blade contour lowermost point with the blade center of rotation) : It was assumed that the extracted lower-most blade geometry contour point in the image plane nearly overlaps with the surface centroid of the propeller.

To effectively drive the UAV to an altitude where using the motion prediction model (8) $(x_p, y_p)$ aligns with the center of the image, visual feedback closed loop setting had to be specified. A visual servoing control loop based on (3) and (4) was developed in the framework of Position-Based Visual Servoing. This approach was chosen as the control implemented in this study aimed at tracking trajectories based on position and orientation error quantities in the Special Euclidean Group *SE(3)*. Utilizing RGB, GPS and IMU, the only unknown quantity to invert each equation in (4) is the longitudinal distance to the detection in camera frame $x_c$. To address this issue, the ratio of the vertical pixel difference between the sought blade point and the principal point to the difference between the vertical world coordinate of the point and the UAV was defined:

$$\begin{aligned} \lambda &= \frac{y_p - v_0}{\Delta z_{des}}, \\ \Delta z_{des} &= z_{i,des} - z_w, \end{aligned} \quad (9)$$

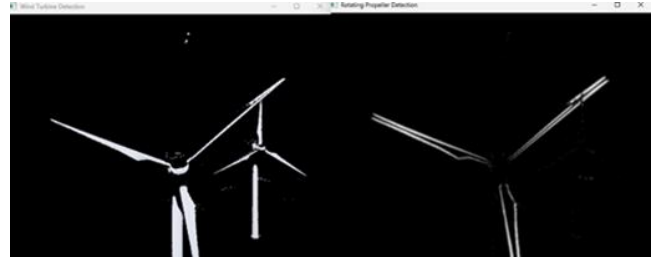

**Fig. 4.** Motion detection. From the specific viewpoint (left image), only the left-most wind turbine is observed to be rotating, detected by the digital filter (right image).

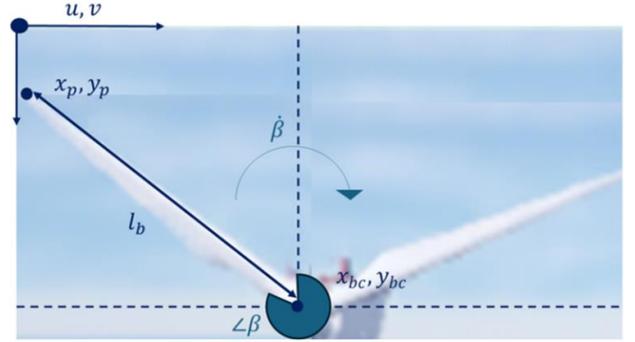

**Fig. 5.** Illustration of the retrieved motion quantities in the image plane.

where $z_{i,des}$ is the vertical position of the blade upper point in the world frame. $\lambda$ was estimated in a data-driven active perception manner. The drone was requested to track vertical position setpoints equally spaced with 1 meter distance from one another. As the UAV was navigating, the pixel difference of the predicted blade upper point location between consecutive frames as well as the difference between consecutive altimeter sensor readings were stacked. In order to avoid an inaccurate estimate due to any transients occurring in the perception system, the termination criterion of the active data collection was:

$$\dot{\hat{\lambda}} = 0, \quad (10)$$

for *m* consecutive nonzero measurement differences. Then, the position command for the UAV to drive the error $e = y_p - v_0$ to zero is:

$$z_{i,des} = z_w + \frac{1}{\lambda}(y_p - v_0) \quad (11)$$

The longitudinal distance to the object of interest is found by plugging (11) into (4) for the calibrated position where (3) holds and by utilizing the related bounding box height. This is achieved in particularly by solving (4) with respect to $x_c$.

### D. Object Tracking

Object tracking module updates the Extended Kalman Filter with 2D vision measurements whilst not relying on very heavy computations. In this study, an initial *YOLOv8* bounding box was used to extract the region of interest and subsequently perform *Canny* edge detection in the same fashion as in *Active Height Measurements*. Following to the extraction of the contour points, the Lucas - Kanade algorithm [23] was integrated. The mean vector of the Lucas – Kanade tracked image points was filtered by calculating the bounding box enclosing the points and implementing a constant velocity Linear Kalman Filter as in [24]. In Fig. 6 the pipeline developed for object tracking is demonstrated.

### B. EKF Localization Module

Let the position vector $x_c$ of the tracked image point in *SE(3)* obey the following kinematics [Grabe et al., 2012] :

$$\dot{x}_c = - \Omega\, x_c - v_c \quad (12)$$

$v_c \in \mathbb{R}^3$ is the vector of relative linear velocities. As the detected target is stationary, it is constituted only of the drone translational velocities expressed in the body/camera frame. $\Omega = \omega^{\wedge}$ is the skew symmetric matrix form of the vector of the body angular rates $\omega \in \mathbb{R}^3$. The measurements' model of the EKF was presented in (4). It should be noted that the mean vector of the Lucas – Kanade tracked features was used for a representative point in the surface of the object of interest. The EKF localization can be extended as well to multiple image features for sparse 3D reconstruction, as proposed also in [14].

### C. Planning and control

To provide a trajectory that has the smoothness property of the $C^2$ class of continuously differentiable functions, a third order polynomial motion timing law was devised.

Given an initial position / velocity in the world frame, a final position / velocity, and initial and final times, the coefficients of the trajectory to be computed need to validate the boundary values as well as 2 intermediate timing law points. Instead of deriving the analytical solution, an approximate solution that minimizes the magnitude of the acceleration over a candidate trajectory was sought:

$$\Gamma = \int_{t_0}^{t_f} \|\ddot{r}(t)\|^2 dt \quad (13)$$

$r_i = (x_i, y_i, z_i)$ is the UAV position vector on the *i*-th trajectory point, expressed in the world frame. The world and body / camera frame follow the North – East – Down (NED) convention.

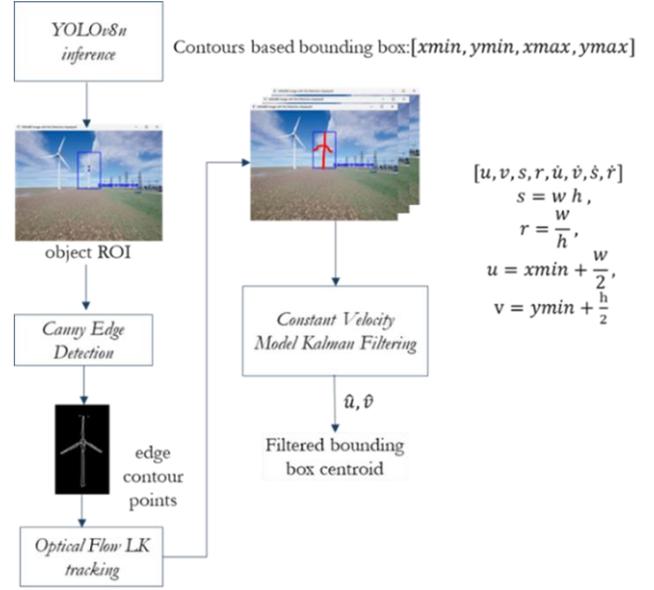

**Fig. 6.** Object Tracking. Canny Edge detection is applied. The features are subsequently tracked by Lucas Kanade Optical Flow algorithm and filtered by a Linear Kalman Filter.

The parametric candidate trajectory is expressed as in [25] :

$$r_{des}(t) = \sum_{i=0}^{3} c_i (t - t_0)^i \quad (14)$$

$$\dot{r}_{des}(t) = c_1 + 2c_2(t - t_0) + 3c_3(t - t_0)^2 \quad (15)$$

$$\ddot{r}_{des}(t) = 2c_2 + 6c_3(t - t_0) \quad (16)$$

The following quadratic form serves as a proxy to (13):

$$\min_{c} c^T \int_{t_0}^{t_f} \begin{bmatrix} 0 \\ 0 \\ 2 \\ 6(t-t_0) \end{bmatrix} [0 \ \ 0 \ \ 2 \ \ 6(t-t_0)] dt\, c \quad (17)$$

The quadratic form (17) was minimized with respect to the coefficients vector $c$, subject to the kinematics in (13)-(15) and the initial and final states. The timing law interval was defined such that the drone exhibits a feasible acceleration between two consecutive via points, considerably below the motor torque and thrust saturation limits.

The quadrotor dynamics considered are [26], [15],[27]:

$$\dot{r} = v \quad (18)$$

$$\dot{v} = \ddot{r} = -\frac{1}{m} R\, T - \frac{1}{m} R\, D\, R^T \operatorname{sgn}(v) \|v\|^2 + g\, z_g \quad (19)$$

$$z_g = [0 \ \ 0 \ \ 1]^T \quad (20)$$

$$\dot{R} = R\, \Omega \quad (21)$$

$$\dot{\omega} = I^{-1}(\tau - \omega \times I\omega) \quad (22)$$

$T$ is the total thrust control vector applied along the vertical body axis, $\tau$ is the body axes' torque control vector and $\mathbf{D}$ is a diagonal matrix of the drag coefficients. $g$ is the gravitational acceleration, $m$ is the mass of the drone and $\mathbf{R}$ is the rotation matrix that converts from the body to world frame. The latter is aligned with the standard definition of the rotation matrix as a synthesis of the 3 elementary rotations: $\varphi, \theta, \psi$ (RPY). This definition has been frequently used in the related literature. $\mathbf{I}$ is the quadrotor inertia tensor. The $12 \times 1$ state vector of the system in (25)-(29) is constituted of the position vector, the linear velocity vector, the Euler angles and the body angular rates. The $4 \times 1$ control input vector is constituted by the body axes' torques and thrust.

The motion control implemented in the current study shares some similarities with the one proposed by Nagami [18]. Their similarities are w.r.t the selection of the same state vector for feedback and the adaptation to the same level of control APIs provided by *AirSim*. The aim of the implemented control law for active perception was to efficiently control the UAV state vector whilst tracking modest accelerations. Thus, a modification of an effective and computationally performant algorithm that minimizes position and orientation error was devised. The difference with respect to the original algorithm proposed by [19] is that the position error is projected onto the tangential direction of the desired trajectory, in addition to projecting it on the normal and binormal directions. This path following controller was used in the *Active Height Measurements* planar active perception phase. Finally, for the hover control in the *Active Height Measurements* climbing stage, a Proportional Integral Derivative (PID) controller for small deviations from hovering state was used.

## IV. RESULTS

A key element in the proposed pipeline is the accurate object detection that provides the foundations for 3D information extraction and subsequent tracking. *YOLOv8* model was fine-tuned on wind turbine and electric tower images. To lower the computational burden whenever unnecessary, two individual detection models were trained, one for each object class. For the training procedure, 200 images were collected for each class. As regards the wind turbine images, the 95% were collected from the *AirSim* environment provided with [28]. The remaining 5% was constituted of real-world UAV photos collected from [29]. The electric tower images were exclusively from *AirSim*. The entire annotated dataset is available at [30].

The collected images were further divided to a **70-15-15%** training-validation-test set split. Training was facilitated by the *Ultralytics* Python package. The training hyperparameters used as well as the weights of the Binary Cross Entropy loss for object classification and Distribution Focal and Complete Intersection Over Union (IOU) losses for bounding box predictions are available in [31]. Fig. 7 shows the object detection capabilities of our fine-tuned *YOLOv8* on two images not seen during training and validation. Fig. 8 illustrates the precision - recall curves for the validation dataset.

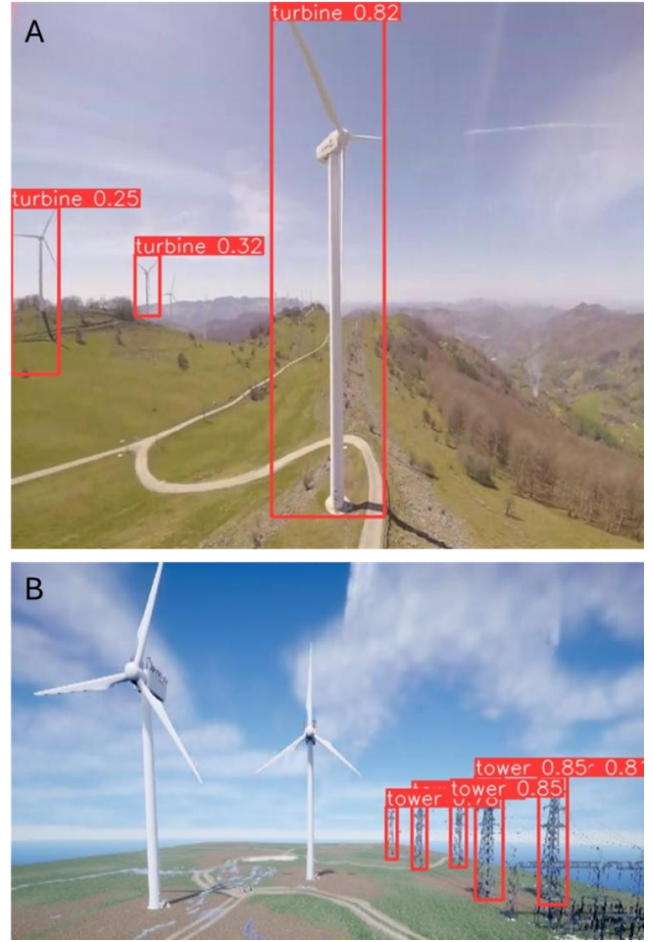

**Fig. 7.** Example fine-tuned *YOLOv8* inference on an unseen image where wind turbine detections were sought (A) and on an image seeking electric towers (B).

The integration of *ActivePerceptionNet* in the active *YOLOv8* inference loop yielded detections with object confidence approaching the peak of the trained detectors' capacity. In Fig. 9 a comparison of the standard vs. active YOLOv8 inference with respect to object confidence is visualized. As it is highlighted with the red indicators in Fig. 8, the wind turbine confidence score is approximating the peak value of **0.974** with the integration of *ActivePerceptionNet*. In the standard YOLOv8 inference, observing the wind turbine yields a nearly periodical object confidence fluctuating between **0.90** and **0.975**. Moreover, from the collected data, the periodicity of the near uniform circular motion of the propeller was retrieved in order to initialize the velocity in the motion model (8). In Fig. 10, a snapshot of the blade contour tracking process using (8) is depicted.

Next, the efficiency of *Active Height Measurements* on estimating the height of a textured stationary object, in this particular case an electric tower, is demonstrated in Fig. 11.

The electric tower assigned with the lowest object confidence upon *YOLOv8* inference from the initial calibrated position was considered. Fig. 10 depicts the progress of the algorithm at the initial viewpoint, the intermediate point where planar motion was terminated upon meeting the object size criterion, and the viewpoint from where the tower height estimate was retrieved. Table 1 contains the collective results of **10** simulation runs for each object class. The ground truth height for the simulated wind turbine was **77.48** meters, measured from ground level to the top vertex of the propeller when oriented as the 'Mercedes-Benz' logo. The ground truth height for the electric towers was **31.88** meters. The average estimated height for the wind turbine was **77.52** meters with a standard deviation of **0.04** whilst the average estimated height was **30.98** meters with a standard deviation of **0.01** respectively. Additionally, the Root Mean Squared Error (RMSE) metric was used for evaluation:

$$RMSE = \sum_{i=1}^{N} \sqrt{\frac{\|z_{w,object}(i) - \hat{z}_{w,object}(i)\|^2}{N}} \qquad (24)$$

As illustrated on Table 1, the lowest RMSE value was achieved for the case of the height estimation of wind turbines, namely **0.06** meters.

To demonstrate the efficacy of the Position-Based Visual Servoing control law in (9)-(11) Fig. 12 illustrates the vertical pixel error of the uppermost blade upper vertex w.r.t the vertical coordinate of the image principal point. The UAV sequentially translated in the vertical direction by 1-meter per iteration to collect data for $\lambda$ estimation whilst the motion model (8) was utilized to predict blade upper vertex position. Active parameter estimation endured **21** seconds approximately. Upon the termination of parameter estimation, the vertical pixel error $e$ approached zero in **5** seconds, using the Position-Based Visual Servoing control law.

Finally, in Fig. 13 the enhancement on the distance estimate to the object of interest in the camera frame is showcased. As it can been observed, the initial depth estimate, which was based on the height estimate, is slightly different than the ground truth value which is not known by the perception system. Although the height estimate error was observed to be at sub-meter error level (Table 1), the bounding box does not utterly match the wind turbine height in pixels. This is due to the fact that the bounding box generated at the initial, calibrated position, does not compensate for the propeller orientation previously presented – the 'Mercedes Benz'. The longitudinal distance error thus was further reduced by the EKF. smooth trajectory was planned towards the imperfect estimate of the point 3D location. Following a trajectory towards the estimated point for **15** seconds, subject to (14) – (22), the 3D estimate was improved through the EKF.

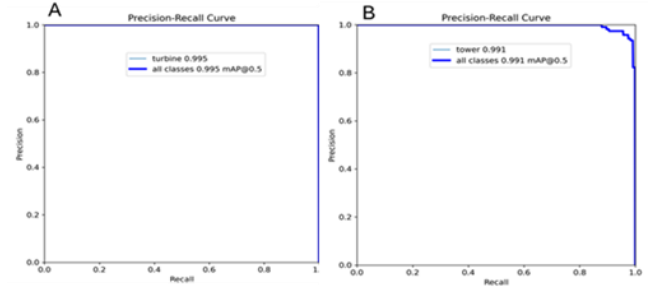

**Fig. 8.** Precision-Recall curves on wind turbines validation set (A) and electric towers validation set (B).

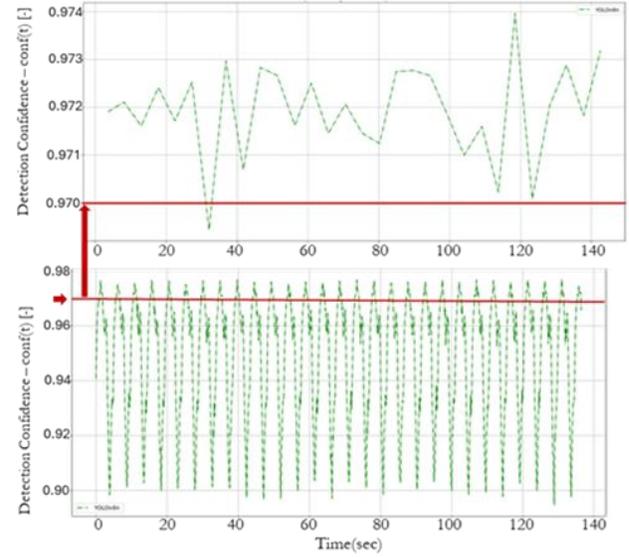

**Fig. 9.** Comparison of standard (bottom) vs active (top) *YOLOv8* inference w.r.t confidence of the wind turbine being enclosed in the predicted bounding box.

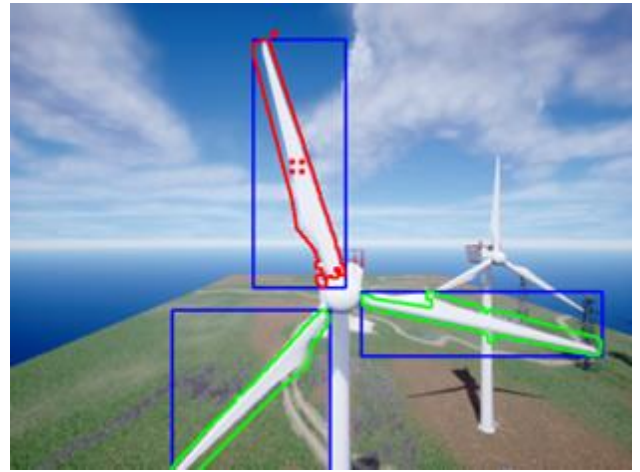

**Fig. 10.** Blade tracking. The blade at the uppermost location in the initial frame is tracked (red contour) based on the motion model (8) that approximates the top vertex position (red dot).

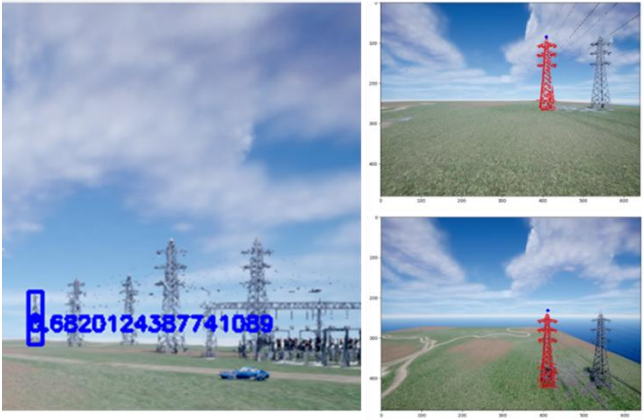

**Fig. 11.** Estimation on a tower (left). The intermediate viewpoint (upper right) corresponds to a size threshold proportional to the ratio of tower to wind turbine sizes. The final viewpoint (lower right) depicts the final measurement.

TABLE I
HEIGHT ESTIMATION RESULTS

| Object class | *Wind Turbine* | *Electric Tower* |
|---|---|---|
| mean ± std | *77.52 ± 0.04* | *30.98 ± 0.01* |
| RMSE ($N = 10$) | *0.06* | *0.89* |

## V. CONCLUSION

An active perception system has been presented in this study, relying on cost-efficient sensors for vision-based navigation, in particular GPS, IMU and RGB sensors. The system consists of the *Object Tracking*, *EKF based localization* and *Planning and Control* modules. Two additional key model pipelines were plugged in the perception system to enhance the initial estimations on wind turbine and electric tower detections. The first one was developed around *ActivePerceptionNet*, a deep convolutional network that learned the mapping between the latent periodicity in the wind turbine appearance due to rotational motion and object confidence. By learning to predict the next confidence peak and actively repeating object detection, this part of the study aimed at answering the "When to look?" question. The other is *Active Height Measurements*, a pipeline that combines performant vision methodologies with path following and hover control. By actively navigating the UAV towards the uppermost point of the object structure, this part of the study aimed at answering the "Where to look?" and the "Why" of the aerial robot's actions. Central to all the aforementioned pipelines is the *YOLOv8* object detection model, fined tuned on wind turbine and electric tower images. The object detector integration aims at answering the "What to look for?" question with an increased confidence. Given that an estimate of the detected object's height and distance is retrieved, the object tracking procedure is initiated. Optical flow based and IMU information are fused in the EKF based localization module. The overall perception – control loop is efficient by means of computational resources and capable of real-time execution.

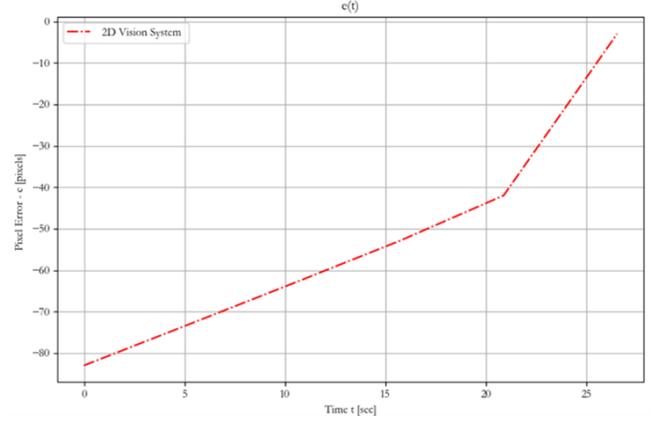

**Fig. 12.** Position – Based Visual Servoing. The UAV reached the propeller blade upper vertex upon $\lambda$ estimation, lasting 21 seconds. The pixel error converged to zero at 26 seconds.

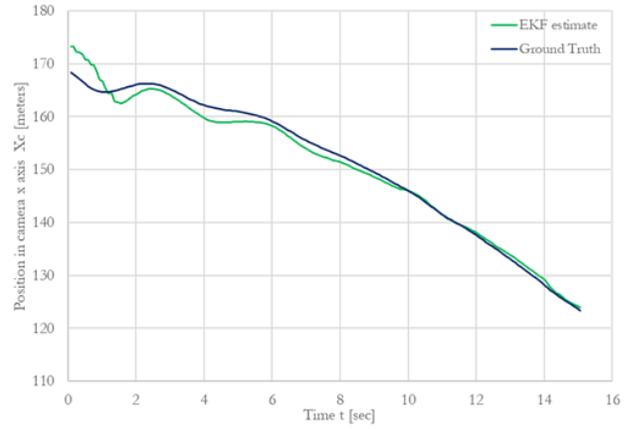

**Fig. 13.** Camera longitudinal distance to the object surface point of interest. The blue trace corresponds to the ground truth and the green trace to the EKF prediction.

A variety of simulation results were showcased, residing in the spectrum of robot vision for object detection, motion detection, visual servoing and state estimation. The results illustrate the potential of the proposed active perception system in a photorealistic environment. As the developed software was tested in real-time simulations, the potential of the proposed system in a real UAV setup can be seen to a certain degree. Finally, it is noted that the results of the simulation experiments showed efficient object height estimation. It particular, a sub-meter error level was achieved. An increased detection confidence near the peak of the deep model's capability was achieved via active inference. The aforementioned procedures were followed by smooth trajectory tracking and simultaneous Kalman filtering, further improving the UAV localization.

**Konstantinos Gounis** received a Diploma /Integrated Master degree in Production and Management Engineering from Democritus University of Thrace, Greece in 2013. He worked in the Automotive Industry of UK.
He is currently a researcher at the Aristotle University of Thessaloniki, Greece. His research interests include Motion Control, Computer Vision and Statistical Machine Learning Algorithms.

**Nikolaos Passalis** (Member, IEEE) received the B.Sc. in Informatics in 2013 and the Ph.D. degree in Informatics in 2018. He is currently a postdoctoral researcher at the School of Informatics, Aristotle University of Thessaloniki. His current research interests include Deep Learning and Data Analysis.

**Anastasios Tefas** (Member, IEEE) received the B.Sc. in Informatics in 1997 and the Ph.D. degree in Informatics in 2002, both from the Aristotle University of Thessaloniki, Greece. Since 2022 he has been a Professor at the School of Informatics, Aristotle University of Thessaloniki. His current research interests include computational intelligence, deep learning, pattern recognition, machine learning, digital signal and image analysis and retrieval, computer vision and robotics.